# A Perspective on the Role of Human Behaviors in Software Development: Voice and Silence*

Mary Sánchez-Gordón
*Department of Computer Science and Communication*
Østfold University College
Halden, Norway
mary.sanchez-gordon@hiof.no

Ricardo Colomo-Palacios
*Escuela Técnica Superior de Ingenieros Informáticos*
Universidad Politécnica de Madrid
Madrid, Spain
ricardo.colomo@upm.es

Muhammad Azeem Akbar
*Software Engineering Department and Systems*
LUT University
Lappeenranta, Finland
azeem.akbar@lut.fi

Monica Kristiansen Holone
*Department of Computer Science and Communication*
Østfold University College
Halden, Norway
monica.kristiansen@hiof.no

*Abstract*— Context: Most software companies strive to have high-performing teams and mitigate withdrawal behaviors like being present but unproductive. In this context, psychological safety and developers' perceived impact are suggested as potential drivers of voice and silence behaviors. However, understanding these social aspects of software development entails the incorporation of social science theories. Objective: This study aims to empirically demonstrate whether such a new theory about voice and silence at work actually applies to the software development context. Method: We plan to use a survey questionnaire design. This study will collect data from software development teams and analyze the result using structural equation modeling (SEM) technique. It can contribute to extent of the body of knowledge about the topic.

## I. Introduction

Understanding human factors is important in the context of the practice of software engineering (SE) [1]. Researchers and practitioners alike have increasingly started to recognize that human factors influence the role of the software developer [2]–[4], team performance [5], [6] and productivity [7] as well as their impact on the software development approach [8], [9] and product characteristics like software quality [10], [11] and security [12]. Beyond software developer' characteristics —e.g., technical skills, soft skills, and personality— software practitioners produce software in teams [13], teams of teams [14], or distributed teams [15]. Although agile approaches in software work are meant to simplify teams [16], complex development scenarios characterized by a mix of inshore and offshore teams are leading to new studies on teams and teamwork.

Software development requires a high level of teamwork and communication. According to [17], communication is the process by which individuals exchange thoughts, feelings or information. Communication is also reported recurrently in the literature as a big challenge for software teams [17]–[20]. In current scenarios, software practitioners across all roles must incessantly communicate with teammates as well as with project stakeholders, including users, analysts, suppliers, customers, and business partners [21]. However, software practitioners speaking up on opportunities or concerns to teammates or upper management are referred to as engaging in voice behavior according to [22] based on the review conducted by [23]. Voice behavior is distinct from other types of workplace communication [24] and is considered a valuable asset for enterprises [20], [25].

In work settings, employees' silence and voice are two separate concepts. Employees observe inefficient work procedures, aspects to improve, or inappropriate work behaviors on a regular basis [25]. Voice can be defined as chances for personnel to have a say and possibly impact organizational affairs relating to matters that influence their work and the interests of managers and owners [26]. According to [27], voice is informal, discretionary, and change-oriented. Voice benefits for organizations include improved generation of ideas, better adaptation to changes, or enhanced problem identification [25]. Voice is essential for ensuring that team members' expertise and competence are used properly inside teams [28]. Research has enriched our understanding of how voice may be beneficial for error prevention, organizational functioning, and innovation [29] and how voice affects the voicers and their teams and supervisors [24]. On the employee side, voice mechanisms increase alignment between organizational and personal goals and enhance productivity, commitment and engagement [30], [31]. Through resources like voicing ideas, feedback provision, and knowledge sharing, supportive practices and relationships at work foster psychological safety and influence positive work outcomes such as learning, performance, innovation, and creativity [32]. Although voicing new ideas potentially benefits an organization, it carries certain risks for employees [25]. For instance, the established way of doing things might be challenged by the voicing of new ideas that go against the vested interests of others team members.

On the other hand, silence may reveal circumstances where employees either do not have chances to voice or do not use them for several reasons [33]. For instance, silence is reported as adverse because employees not communicating their ideas and concerns could not only be harmful to psychological and physical health [25] but may also damage organizational interests [34]. So far, much of the research on silence has investigated it in isolation [24] and is scarce [24], [25]. For employees, silence may cause feelings of anger and/or fear, reduce job satisfaction and creativity, and increase emotional exhaustion, deviance, and burnout [24]. As a result of the importance of voice and silence, the decision-making process of employee voice and silence behaviors is worth scholarly inquiry [25].

In the software engineering (SE) literature, it seems that voice and silence are under-explored constructs. There is a scenario-based experiment [35] to explore three factors that contribute towards team member silence behaviors in a

---

distributed software project: the individual's level of experience, the role of the offending team member, and the individual's responsibility to report. Findings reveal that some of the factors from the silence literature may not be relevant in the context under study. Therefore, further research is needed to examine the phenomena of employee silence. Another experiment [36] investigates how voice and related dimensions of group process (group work and discussion quality) are affected by group size and social presence. Findings show that social presence impacts instrumental and value-expressive voice, as well as related outcomes of group process. Other studies have not focused on or measured voice or silence but have mentioned them. For instance, Greiler et al. [37] identified that developers are more eager to voice and tackle problems to continuously improve developer experience on teams with good psychological safety. In addition, a recent systematic mapping of burnout in SE [38] identified that communication practices within a development team can lead to burnout among software developers. These practices include impolite requests, lack of participation in terms of 'speaking up', openness to criticism, and disagreement among co-workers. Finally, a recent systematic literature review [39] found that whistleblowing —another area of literature related to employee silence— is also an understudied area of SE research.

Despite the large body of work that has been conducted to understand voice and silence behaviors across a wide range of industries, including manufacturing, retail, and healthcare [24], only a few studies exist that have analyzed the effect of those behaviors on software teams. Our long-term goal is to understand the role of voice and silence behaviors in SE context. However, understanding these social aspects of software development entails the incorporation of social science theories to prevent the over-rationalization of core phenomena [40]. Research also highlighted the need for further understanding of whether silence has different predictors than voice [24]. Therefore, this study aims to test a new theory proposed by Sherf et al. [41] about voice and silence at co-located teams, or hybrid software teams (part-remote/part-office arrangements). Empirically demonstrating whether this theory actually applies to SE context can contribute to the body of knowledge and address the need identified by [40] of testing social theories in SE context before use.

In the following section (2) we briefly present the theory proposed by Sherf et al. [41] and its hypotheses. Then, we present the variables (3) and, finally, the participants (4) and execution plan (5) are presented.

## II. HYPOTHESES

This study formulates the following research question: Is it possible to test the theory proposed by Sherf et al. [41] in SE context?

The hypotheses are based on established theories and salient themes from previous studies. Fig. 1 illustrates the theoretical model and hypotheses. The theory hypothesizes that the psychological safety and perceived impact will both be related to team silence (H1 and H3) and voice (H2 and H4) behaviors. On the other hand, silence and voice behaviors will influence psychological withdrawal (H5 and 7) and team performance (H6 and H8).

### A. Behavioral Activation and Behavioral Inhibition Systems

Employee voice is a term that has been utilized in a variety of contexts and disciplines over time. Despite the advances in the voice and silence literature [24], similar mediating judgments, e.g., efficacy judgments, can be found in many studies so that the logic and expected findings would be anticipated in the opposite direction by using the word "silence" instead of "voice". However, this view was challenged by Sherf et al. [41]. These authors posit that voice and silence are different because they reflect two self-regulatory systems. Voice is an approach behavior that reflects the behavioral activation (approach) system (BAS) whereas silence is an avoidance behavior that reflects the behavioral inhibition (or avoidance) system (BIS). They carried out a meta-analysis and a panel study over six months. As a result, they concluded that the frequency of voice and silence is determined by the extent to which these independent systems are triggered. In our study, we rely on that proposed conceptual framework for the independence of voice and silence.

Previous literature argues that the BAS and BIS are two self-regulatory systems that underlie behavior [42]. The BAS regulates appetitive motive and aims to move toward —i.e., approaching, seeking, or achieving— potential rewards or opportunities to the self. On the other hand, the BIS regulates aversive motives and aims to move away from —i.e., avoiding, preventing, or inhibiting— potential harm or punishment to the self. Moreover, different environmental stimuli trigger the BAS and BIS which are mutually independent. There is also evidence that BAS and BIS are connected to emotional states, with positive or negative emotions occurring when moving towards positive change or positive stimuli (e.g., rewarded) or away from negative change or negative stimuli (e.g., punished).

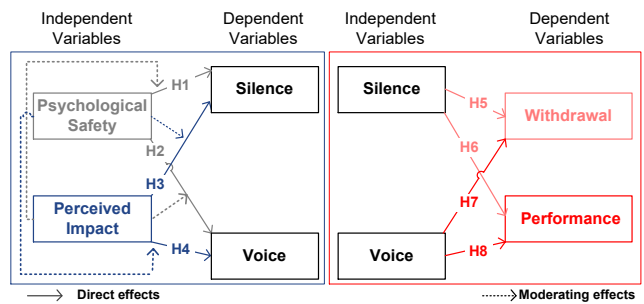

Fig. 1. Theoretical model and the hypotheses

### B. The Influence of Psychological Safety and Perceived on Voice and Silence Behaviours

Sherf et al. [41] also found that "perceived impact and psychological safety differentially predict voice and silence behaviors". Findings showed that voice is more strongly affected by perceived impact (through the BAS activation), but silence is more strongly influenced by psychological safety (by the BIS activation). Our study investigates psychological safety since it is considered as a valuable resource [32]. Software industry studies identified psychological safety as a top driver of team and business performance [37]. Additionally, we will investigate a motivational state called perceived impact (influence) which has been found to affect teams in the voice and silence literature but has been overcome by SE researchers even though "locus of control" has been studied. Perceived impact differs from a locus of control in that an individual's sense of impact is influenced by the work context whereas internal locus of control is a personality trait that persists across situations [43].

Employees' perceptions of impact seem a proper environmental cue for the BAS since perceived impact is the degree to which employees believe that they can influence outcomes at their work [41]. Perceived impact can alert employees about their ability to gain opportunities or rewards by voicing knowledge or information. For instance, the project team member's perception of the supervisor's openness to the information about an underperforming project's flaws has a positive impact on the willingness to voice concerns about that project to their supervisor [44]. However, in practice, the style of leadership is probably less important than if the employee feels that their manager is supportive, sets a positive example, and creates an environment where speaking up is safe [24]. Employees, who perceive greater impact, sense higher levels of control over their work environment and develop a greater sense of responsibility to engage in change-oriented behaviors like voice [45]. High perceived impact levels would increase voice with the objective of achieving desired work outcomes. However, if the environment creates a sense of learned helplessness in which no opportunities to achieve gains or rewards exist, the BAS is less likely to be triggered, leading to low infrequent levels of voice [41]. Approach behaviors like voice are not triggered when there is a low perceived impact because rewards or opportunities are either absent or not achievable. It suggests that silence, as an inhibition-oriented behavior, will be less strongly influenced by perceived impact because it is less associated with BIS states triggered by the presence of risk —fear, anxiety and vigilance— or by the absence of risk — security, relief and calmness [41].

Psychological safety seems a proper environmental cue for the BIS since psychological safety refers to the degree to which behaviors might have negative interpersonal outcomes [24], [46]. Psychological safety is a key factor of high-quality communication, decision making and trust that plays a vital role within workplace teams [32], [47]. Psychological safety facilitates interpersonally risky behaviors like speaking up and voicing [25], [32], [47]. According to [32], previous work on enhancing psychological safety also suggests that employees feel safer sharing ideas and speaking out when the higher the status of the team or employee is. In particular, experienced employees with high work positions are more likely to believe that speaking up will bring them positive outcomes [25]. However, although employees think their contribution would be valuable to the team, they sometimes choose silence over voice when there is a lack of psychological safety [29]. Lack of psychological safety would increase employee silence with the objective of avoiding potential threats, risks or dangers in the work setting. However, if the environment is psychologically safe, the BIS is less likely to be triggered, leading to low or infrequent levels of silence [41]. Avoidance behaviors like silence are not triggered when there is high psychological safety because there will be no negative stimuli (potential harm or punishment). It suggests that voice, as an approach-oriented behavior, will be less strongly influenced by psychological safety because it is less associated with BAS states like hope, enthusiasm, or eagerness [41].

Given that psychological safety and perceived impact can take place simultaneously, e.g., [41], [45], we include these predictors as control variables in our hypotheses — as recommended by [48]— and formulate the following:

***Hypotheses H1 and H2*** *Psychological safety in software development teams relates more strongly to their silence than to their voice when controlling for the effect of team members' perceived impact.*

***Hypotheses H3 and H4****: Software development team members' perceived impact relates more strongly to their voice than to their silence when controlling for the effect of psychological safety.*

### C. Relationships of Voice and Silence with Withdrawal and Performance

Sherf et al. [41] also found that silence relates to burnout more strongly than voice and suggested further research to analyze the differential relationships of voice and silence with performance. This study will examine psychological safety and burnout because both have been recognized as relevant factors that affect software development. A recent study [37] identified "speaking up" as a strategy to improve the developer experience and "no longer speaking up" and "reducing engagement" about the problems as coping mechanisms. The last implies that practitioners still performed only absolutely necessary duties which means "withholding work efforts" a characteristic of withdrawal behavior. Withdrawal and exhaustion are two dimensions that characterize burnout [49].

Prior research [41] suggests that frequent BIS activation can have a negative impact on well-being costs by increasing aversive states like strain, stress, or fatigue. A longitudinal study [50] found reciprocal effects between silence, abusive supervision, and fear. Abusive supervision increased fear and led to more defensive silence, which in turn resulted in more abusive supervision. These effects were intensified by lower assertiveness and higher climate-of-fear perceptions. Moreover, a four-wave longitudinal study [51] found reciprocal effects between burnout and silence. Two imposed forms of silence (acquiescent and quiescent) were related to increased levels of subsequent burnout, which in turn related to increased silence. Likewise, a survey study [52] revealed that silence leads to burnout which results in weakening performance, an increase in: withdrawal behaviors, and turnover intentions. Through the mediation of job burnout, silence has a more pronounced negative effect on employee performance and a positive impact on turnover intentions. Findings of these studies show that silence has a consistently positive relationship with burnout. From an approach-avoidance perspective, these findings can be understood as a movement away from negative stimuli.

On the other hand, voice effects on occupational well-being have been not well understood [24], [53]. Engaging in voice can potentially damage employees' well-being when leading to harmful consequences, e.g., damaged relationships with co-workers and superiors, or loss of career opportunities [53]. It also foster situations where employees talk about potentially uncomfortable issues and defend alternative points of view, making voice a highly resource-demanding behavior. Moreover, a lack of understanding of how voice may impact well-being is critical since low well-being can result in deterioration of health, withdrawal, and higher turnover [53]. Depending on how (un) favourably others perceive voice, it may either increases or reduces withdrawal and exhaustion [25]. However, research identified that approach motives and sensitivity to BAS have no effect on outcomes similar to burnout, e.g., strain [54]. As a result of this uncertainty, we suggest that voice's relationship with burnout should be weaker than the relationship between silence and burnout. Hence, we propose the following hypothesis:

***Hypotheses H5 and H7**: Software development team members' silence presents a stronger association with withdrawal compared to their voices.*

Research on performance is related to voice (a response to BAS). As BAS is related to obtaining and strive for opportunities and rewards, engaging in voice may affect performance by signaling an employee's prosocial contributions to the team [27]. Recently, a longitudinal approach found that voice more positively impacts team effectiveness when teams face discontinuous change [28]. Particularly, prohibitive voice reduced performance losses by enabling error management at the discontinuous change phase and promotive voice enhanced performance gains by enabling process innovation at the recovery phase. Moreover, a meta-analysis [55] identified that the relationship between high-performance managerial practices and job performance is mediated by voice. Research also suggests that group and organizational effectiveness perform suffers when there is a high level of silence and performance is better when employees share their ideas and concerns [27]. As mentioned before, BIS regulates aversive motives and can lead to strain. Silence may have a negative impact on performance by consuming employees' energy and eventually lowering feelings of personal accomplishment [27], [51]. Hence, we propose the following hypotheses:

***Hypothesis H6 and H8**: Software development team members' voice presents a stronger association with team performance compared to their silence*

## III. VARIABLES

This study aims to investigate human behavior and the correlation between team performance and burnout in software development projects. Human behavior will be measured using individuals' voice and silence, perceived impact, and psychological safety. In the survey, we ask respondents to think about their last month at work and indicate how much they agreed with these sentences. Table 1 lists the items used to compile the six variables.

**Perceived impact (Pi)** is the extent to which individuals believe that they can influence outcomes at their work [41]. To capture respondents' perceived impact on their teams, we used a three-item measure (1 = "strongly disagree" to 5 = "strongly agree") [43].

**Team performance (Tp)** is the extent to which a team can meet established objectives [56]. Team members were asked to rate their team performance in terms of efficiency and quality. To assess team performance, we use six items (1 = "strongly disagree" to 5 = "strongly agree" defined in [5] based on the items suggested in [56].

**Psychological safety (Ps)** is defined as "a shared belief held by members of a team that the team is safe for interpersonal risk-taking" [57]. For the BIS is relevant and highly salient that perceptions about the possibility that one's communication will result in a negative outcomes such as discomfort or rejection [41]. To capture individuals' experience of psychological safety in their team, we used a seven-item measure developed in [57]. Items were each measured on a seven-point Likert scale (1 = "strongly disagree" to 7 = "strongly agree").

**Voice (Vo)** is the expression of concerns and ideas with the goal of influencing others in order to induce improvements or changes [41]. Workplace response to BAS regulation is voicing issues to obtain successes, rewards, or opportunities [41]. In the last decade, according to [24], there has been increasing use of the measure proposed by Liang et al. [46]. To capture respondents' voice, we utilized a promotive (three items) and prohibitive (three items) voice measure (1 = "Never" to 7 = "Daily") from Liang et al. [46] and adapted by [41].

**Silence (Si)** is the suppression of concerns and ideas with the goal of preventing or avoiding communication that could cause harm to oneself [41]. Workplace response to BIS regulation is withholding issues to avoid failures, risks, or punishments. The measures proposed by Tangirala and Ramanujam [45] and Detert and Edmondson [29] are the most popular according to [24]. To capture respondents' silence, we utilize a three-item silence scales (1 = "Never" to 7 = "Daily") adapted from those measures by [41].

**Burnout (Bo).** As withdrawal more so than exhaustion can result in an immediate negative impact on individuals' and organizations' performance, this study is focused on withdrawal. In addition, exhaustion is often subsumed by withdrawal since individuals deal with exhaustion by using it as a coping mechanism [58]. To assess psychological withdrawal, we use eight items from [59]. All items are scored on a 7-point Likert scale (1 = "never" to 7 = "daily").

TABLE I. THE TABLE SHOWS THE ITEMS USED TO COMPILE THE SIX VARIABLES: PERCEIVED IMPACT (PI), TEAM PERFORMANCE (TP), PSYCHOLOGICAL SAFETY (PS), VOICE (CV), SILENCE (CS), AND BURNOUT (BO).

| ID | Item |
|---|---|
| Pi1 | My impact on what happens in my team is large |
| Pi2 | I have a great deal of control over what happens in my team |
| Pi3 | I have significant influence over what happens in my team |
| Tp1 | This team produces quality work |
| Tp2 | My team is productive |
| Tp3 | This team delivers according to schedule |
| Tp4 | This team delivers according to budget |
| Tp5 | My team communicates efficiently with others |
| Tp6 | The team could become more efficient [a] |
| Ps1 | If I make a mistake on this team, it is often held against me [a] |
| Ps2 | Members of my team are able to bring up problems and tough issues |
| Ps3 | People on my team sometimes reject others for being different [a] |
| Ps4 | It is safe to take a risk around here |
| Ps5 | It is difficult to ask other members of my team for help [a] |
| Ps6 | No one on my team would deliberately act in a way that undermines my efforts |
| Ps7 | Working with my team members, my unique skills and talents are valued and utilized |
| Vo1 | I proactively gave suggestions for issues that may influence my team |
| Vo2 | I proactively voiced constructive suggestions which are beneficial to my team |
| Vo3 | I made suggestions on how to improve my team's working procedures |
| Vo4 | I advised against undesirable behaviors that would hamper my team's job performance |
| Vo5 | I spoke up honestly about problems that might cause serious loss to the work team, even when/though dissenting opinions exist |
| Vo6 | I pointed out problems when they appeared in my team, even if that would hamper relationships with other colleagues |
| Si1 | I kept quiet and did not make recommendations about how to fix work-related problems |
| Si2 | I kept ideas about how to improve work practices to myself |
| Si3 | I choose not to speak up with ideas for new or more effective work practices |

| ID  | Item |
|-----|------|
| Bo1 | I think of being absent |
| Bo2 | I chat with co-workers about nonwork topics |
| Bo3 | I leave my work station/area for unnecessary reasons |
| Bo4 | I daydreaming |
| Bo5 | I spend work time on personal matters |
| Bo6 | I put less effort into my work than I should have |
| Bo7 | I think of leaving my current job |
| Bo8 | I let others do my work |

[a.] When creating the index variable, the answer is to be inverted (Reverse scale)

**Control Variables.** We included two control variables in the model: age, experience, and gender identity. Age is defined as a category that includes: 18-25, 26-35, 36-45, 46-55, 56-65, 66+ years. In addition, participants who feel uncomfortable due to this question can select the option "prefer not to say". Age diversity contributes to more effective teamwork according to a recent study conducted by Verwijs and Russo [60]. Experience is measured in years, e.g, practitioners having more experienced may be more familiar with practices that can enhance their confidence to speak up [45]. In the case of gender identity, the survey includes the following options: man, woman, gender variant/non-conforming/non-binary, prefer to self-describe, and prefer not to say. We will expect that most respondents will identify as a "man", so all other respondents will be grouped as "gender-minority", except those who select "prefer not to say". Previous studies [61], [62] suggest a positive relationship between gender diversity and team performance.

## IV. PARTICIPANTS

Given that we are mainly interested in software development teams, the questionnaire was open to all kinds of software professionals, including designers, project managers, quality assurance specialists, architects, and business analysts. However, given the composition of software teams, it is expected that most respondents will identify as developers.

We focus on a convenience sample strategy because we want to collect responses from all team members in order to avoid single-source. Given that response rate is a challenge in sample studies like this, we reduce the time commitment required of respondents but cash incentives for participation will not be offered. The sample will include around 200 software professionals. We plan to contact companies in different countries via email to agree to conduct the survey. The manager of each company will select a sample of teams to participate in this study by considering co-located or hybrid software teams. We will generate a unique identifier for each team (ID-T#), then register the country where the team is based and the number of team members. This information will help us to calculate the response rate of each team and we will retain only those teams with at least half of the participating members. The sample will consist of around 200 software professionals

## V. EXECUTION PLAN

In order to evaluate the theoretical model, we plan to conduct the following set of steps.

**Instrument Design.** We create an anonymous questionnaire survey using Google forms to facilitate self-reported data. Tokens or URL tracking will not be used. Questions are grouped into blocks based on the measures of each construct (also called latent variable). To encourage the participants to truthfully complete the questionnaire, we give information about the study and participation. Moreover, we declare that data will be used for academic research purposes only and inform the participants that the questionnaires are anonymous and that their personal information will be kept confidential. Consequently, instead of reporting organizations' names respondents gave a team identifier number, as part of the questionnaire to identify each team. It is expected that respondents will not be motivated to intentionally misreport since we use these treatments to lessen the impact of social desirability bias.

We will collect the following basic demographic variables: age, gender identity, education, job role and experience. The questionnaire includes six constructs: Psychological safety, Perceived impact, Silence, Voice, Burnout and performance. These constructs refer to abstract concepts that cannot be directly measured [63], therefore we used validated measures as much as possible to improve construct validity. By contrast, direct measurements like years of experience are assumed to have inherent validity because are all directly measurable. However, validation of latent variables is required to ensure that they measure the right properties.

**Pilot test.** We ask for feedback from five colleagues: two SE academics and two software developers. Pilot participants made a few comments on the questionnaire structure, and on the face and content validity of the scales. Based on this feedback we made some changes, moving the Bo scale closer to the end and adding an open-response question. In this way, we assessed content validity.

**Ethical considerations.** We checked the design study to ensure that it will be conducted according to the institutional guidelines of each involved university, including the informed consent document that will be provided to participants. We also created a Data Management Plan (DMP) to define how research data will be processed (storage and accessibility) during and after the study is completed. For instance, we will aggregate answers/data before sharing and only one researcher will manage the contact information of the companies, project managers and the ID-team created. Before starting this study, we will also submit it —including the informed consent, questionnaire, and list of variables— to the competent authorities of each involved university to ensure that it is in compliance with their local and regional legislations and norms.

**Data collection.** The project managers receive an email with instructions and the unique identifier for each team (ID-T#). In turn, they will send an email to the selected team and then, participants will access the survey online via a provided URL. Finally, the project manager will send us confirmation about participation.

**Data Cleaning.** We will delete the consent form confirmation field since respondents could not continue without checking these boxes (the response is always "YES"). Moreover, we will remove participants who do not type the ID team. However we do not plan to remove other outliers unless we find specific reasons to believe the data is not valid.

We will recode the raw data into a common quantitative coding scheme. for example, from 1 for "strongly disagree" to 7 for "strongly agree". The recoding instructions will be included in our replication package. Additionally, we add a field for the country team an other by each reversed indicator to calculate the score using the formula: reverse scale(x) =

max(x) + 1 - x. For example, if the score is 1 and max(x) is 7 because the Likert scale goes up to 7, we take 7 + 1 = 8, and subtract our score from that 8 - 1 = 7.

**Descriptive Statistics.** We plan to present descriptive statistics for the study variables including the demographic variables by reporting means, modes and standard deviations of each one. Then, we will assess potential covariance and relationships between variables understudy.

**Validity Analysis.** We plan to evaluate construct validity using established guidelines by Ralph and Tempero [63]. We will assess convergent and discriminant validity, i.e. *"Do all indicators load on the correct construct?"*. We will conduct two tests: Bartlett's test of sphericity on all constructs (P values < .05) and Kaiser-Meyer-Olkin (KMO) measure of sampling adequacy. After checking possible issues —the correlations between indicators of the same/different constructs— and solving the problems, predictive validity will be evaluated by testing the proposed hypotheses. We will estimate values for the constructs from reflective indicators by using reliability testing (Cronbach's Alpha). Then, we will use inferential statistics to test the expected relationships.

Responding to a single instrument (the survey) regarding independent and dependent variables at the same time can lead to artifactual covariance between these variables. Researchers have suggested this may lead to a systematic measurement error known as Common Method Bias (CMB) (or Common Method Variance) [64]. Our theoretical model also includes two mediators, which serve as both exogenous and endogenous variables. Therefore, we conducted Harman's single-factor test. Using this procedure, all variables in the study are loaded onto a single factor in an exploratory factor analysis which must be less than 50% to verify that CMB is not a significant concern [64]. CMB may be present if a single factor emerges, or the first factor accounts for most of the variance.

**Structural Equation Modeling.** The proposed relationships will test using structural equation modeling (SEM). SEM represents a combination of two statistical methods: CFA (Confirmatory Factor Analysis) and path analysis. CFA has an objective to estimate the latent variables. To design a SEM, a *measurement model* is defined by mapping each *reflective indicator* into its corresponding construct. For instance, each one of the three items comprising the PI scale is modeled as a reflective indicator of perceived impact. Moreover, we will model voice as a higher-order indicator of promotive and prohibitive voice.

CFA is used to estimate the latent construct as the shared variance of its respective indicators. The CFA should converge and all of the indicators should load well on their constructs. After we will reach confidence in the measurement model, we plan to represent the hypotheses as regressions to construct the structural model. The two control variables will be used as predictors for all latent variables. Their predictive power will be tested one at a time, and they will be included in regression only if the difference is marginally significant (p<0.1). Although a p-value greater than normal is more conservative, we will use it since we are dropping predictors rather than testing hypotheses. To analyze the CFA model with ordinal indicators, we will use the weighted least square mean variance (WLSMV).

The model must present a good fit to the data, and all indicators loaded significantly (P values < .05). We will evaluate model fit by inspecting the following indicators [65]: the Root Mean Square Error of Approximation (RMSEA) should be less than 0.06 and the Standardized Root Mean Square Residual (SRMR) should be less than 0.08 while the Tucker-Lewis Index (TLI) and Comparative Fit Index (CFI) should be at least 0.95. The structural model tests the hypothetical dependencies based on path analysis. Moreover, we will use the following information criteria: Akaike information criterion (AIC), Bayesian (Schwarz) information criterion (BIC), and Hannan-Quinn criterion (HQC). They are model selection tools to compare the models fit to the same data. We will compute the information criteria for each candidate model and we will determine which model yields the minimum value for each criterion. Then, we identify the model that minimizes all criteria.

**Validity Threats.** We cannot claim any causality because this research design is a sample study representing correlations. Given the cross-sectional nature of the data, we cannot test causality directly, although the hypotheses imply causation. Overall, causality in the voice and silence literature is still an issue, as cross-sectional surveys have been the predominant empirical approach [24]. The sample size is another major validity threat. However, we plan to draw from multiple sources (companies and countries) to increases sample heterogeneity. In addition, it is expected an under-representation of minorities so statistical analyses can be biased. To minimize the recall error, we use in the instrument design a common way of simplifying the task by shortening the reference period (last month).